\documentclass[aps,preprint,showpacs,showkeys, superscriptaddress]{revtex4}
\usepackage{bm}
\usepackage{graphicx}
\usepackage{amsmath, amsthm, amsfonts,amssymb}

\begin{document}

\title{Topological Properties of Stock Networks Based on Random Matrix Theory in Financial Time Series}

\author{Cheoljun Eom}
\email{shunter@pusan.ac.kr} \affiliation{Division of Business
Administration, Pusan National University, Busan 609-735, Korea}

\author{Gabjin Oh}
\affiliation{NCSL, Department of Physics, Pohang University of
Science and Technology, Pohang, Gyeongbuk, 790-784, Korea \& Asia
Pacific Center for Theoretical Physics, Pohang, Gyeongbuk,
790-784, Korea}

\author{Hawoong Jeong}
\affiliation{Department of physics, Korea Advanced Institute and
Science and Technology, Daejeon 305-701, Korea}

\author{Seunghwan Kim}
\affiliation{NCSL, Department of Physics, Pohang University of
Science and Technology, Pohang, Gyeongbuk, 790-784, Korea \& Asia
Pacific Center for Theoretical Physics, Pohang, Gyeongbuk,
790-784, Korea}

\begin{abstract}
We investigated the topological properties of stock networks
through a comparison of the original stock network with the
estimated stock network from the correlation matrix created by the
random matrix theory (RMT). We used individual stocks traded on
the market indices of Korea, Japan, Canada, the USA, Italy, and
the UK. The results are as follows. As the correlation matrix
reflects the more eigenvalue property, the estimated stock network
from the correlation matrix gradually increases the degree of
consistency with the original stock network. Each stock with a
different number of links to other stocks in the original stock
network shows a different response. In particular, the largest
eigenvalue is a significant deterministic factor in terms of the
formation of a stock network.
\end{abstract}

 \maketitle

\section{Introduction}

The financial market has been known to be the representative
complex system which forms a pricing mechanism according to the
interaction between various assets. In order to understand the
pricing mechanism of a financial market, we need to understand the
interaction between assets traded on the market. A correlation
matrix is the representative measurement used to quantify the
interaction between assets in the field of finance. However,
because the correlation matrix reflects the dynamic properties of
financial markets which are formed by the process of creation,
growth, and decline of various assets, the analysis and
classification of the significant interactions between assets from
the correlation matrix is very difficult to undertake. Theories
and methods created to improve this problem have been introduced
in the past. Among these, the random matrix theory (RMT), which is
able to remove random properties from the correlation matrix, was
introduced as well as applied to the field of finance [1-2]. In
addition, a stock network method based on the minimal spanning
tree (MST) method, which extracts the significant information from
the correlation matrix in viewpoint of the classification of the
connected relationship between stocks, was introduced and applied
to the field of financial research [3-4].

According to correlation matrixes having various properties
created from the RMT method and applied to stock networks, we
empirically investigated the topological properties of stock
networks. Some previous researchers have investigated the
topological properties of stock networks using created data from a
pricing model, widely acknowledged in financial literature. Major
pricing models include the one-factor model [5], three-factors
model [6], and multi-factors model [7]. The purpose of the
previous studies was to discover the possible deterministic
factors which have significantly influenced the formation of a
stock network [8-13]. Stock network methods based on the MST
method use the correlation matrix as an input variable. Therefore,
after creating a correlation matrix having various eigenvalue
properties calculated by the RMT method, we exhibit the stock
networks using those as an input variable. We investigated the
topological properties of stock networks through a comparison of
the original stock networks with the estimated stock networks. We
used the individual stocks traded on the representative stock
market indices of Korea, Japan, Canada, the USA, Italy, and the
UK, respectively. We discovered that as the correlation matrix
reflects the more eigenvalue property, the estimated stock network
from the correlation matrix gradually increases the degree of
consistency with the original stock network. The interesting point
is that each stock with a different number of links to other
stocks in the original stock network shows a different response.
When we did not include the property of the largest eigenvalue in
the correlation matrix estimated by RMT, even though it reflected
all eigenvalue properties except the largest eigenvalue, the
estimated stock network could not adequately explain the formation
of an original stock network. These results suggest that the
largest eigenvalue is a significant deterministic factor in terms
of the formation of a stock network.

In the next section, we describe the data and methods of the
verification process used in this paper. In section III, we
present the results obtained according to our research aims.
Finally, we summarize the findings and conclusions of the study.

\section{DATA and METHODS}

\subsection{Data}

To investigate the properties of stock networks by using the RMT
method, we used the individual stocks traded on the representative
stock market indices of Korea, Japan, Canada, the USA, Italy, and
the UK, respectively. That is, we used the daily prices of 127
stocks in the KOSPI 200 index of the Korean stock market, 202
stocks in the Nikkei 225 of Japan, 118 stocks in the TSX of
Canada, 378 stocks in the S\&P 500 of the USA, 111 stocks in the
Milan Comit of Italy, and 69 stocks in the FTSE 100 of the UK. The
individual stocks that had daily prices for the last 15 years,
from January 1992 to December 2006, were selected from each
country. The returns, $R_t$, are calculated by the logarithmic
change of the price, $R_t=ln(P_{t})-ln(P_{t-1})$, where $P_{t}$ is
the stock price at day $t$.

\subsection{Methods}

The test procedure in this paper can be explained by the following
three steps. First, we determined input data that is needed to
create a stock network. That is, we determined the estimation
process of a correlation matrix reflecting various eigenvalue
properties estimated by the RMT method. Second, we created a stock
network using the MST method. Third, we calculated the survivor
ratio to compare stock networks.

In the first step, we examine the estimation process of a
correlation matrix reflecting various eigenvalue properties
estimated by the RMT method. It is not difficult to create a
correlation matrix having various eigenvalue properties as the
number of stocks $N$ using the RMT method. Using the eigenvalue,
$\lambda_{S(i)}$, and eigenvector, $V_{S(i)}$, we estimated the
correlation matrix, $C_{i=1}^{S(i)} = \lambda_{S(i)} . V_{S(i)} .
V_{S(i)}^{T}$, reflecting various eigenvalue properties, $i=1,2,
\cdots, N$. That is, among the eigenvalues estimated by the RMT
method, we repeatedly created a correlation matrix reflecting the
properties of eigenvalues included in each range, $S(i)=1 \sim i$,
from the fixed starting point of the largest eigenvalue,
$\lambda_{i=1}$, to the minimum eigenvalue, $\lambda_{i=N}$. For
example, $C_{i=1}^{S(1)}$is a correlation matrix reflecting only
the largest eigenvalue property, while $C_{i=1}^{S(N)}$ is a
correlation matrix reflecting the eigenvalues within the ranges
from the largest eigenvalue to the smallest eigenvalue. That is,
$C_{i=1}^{S(N)}$ is a correlation matrix reflecting all the
eigenvalues; therefore, this correlation matrix in the last range
is the same as those created using the real stock returns.

In the second step, we examine the creation process of stock
networks by using the MST method. The stock network visually
displays the significant $N-1$ links among all possible links,
$N(N-1)/2$, based on the correlation matrix between stocks. It
does this by using the MST method. We created two-type stock
networks according to our research purposes. The stock networks
are the original stock network, $Net^{O}$, using the correlation
matrix estimated from the actual stock return, $R_{j}$ , and the
estimated stock network, $Net_{i}^E$, using the correlation
matrix, $C_{i=1}^{S(i)}$, created by the RMT method. In order to
create a stock network, the metric distance, $d_{i,j} =
\sqrt{2(1-\rho_{i,j})}$, relates the distance between two stocks
to their correlation coefficient, $\rho_{i,j}$ [14]. In our study,
the MST is the spanning tree of the shortest length using the
Kruskal algorithm [15]. Therefore, it is a graph without a cycle
connecting all nodes with links. The correlation coefficient can
vary between $-1 \leq \rho_{i,j} \leq +1$, while the distance can
vary between $0 \leq d_{i,j} \leq 2$. Here, small values imply
strong correlations between stocks.

In the third step, we calculated the survivor ratio to measure the
degree of consistency between the original stock network and the
estimated stock network by utilizing the RMT method. The survivor
ratio, $r_{s, L \geq k} \equiv \ 1/M \sum_{j=1}^{M} \ [ FQ_j
(Net_{L}^{O} \bigcap Net_{L}^{E})] / [ FQ_j (Net_{L}^{O})]$, is a
ratio of frequency showing that stocks such as
$FQ_{j}[Net_{L}^{O}]$ that are directly linked with a specific
stock, $j$, in the original stock network have the same links with
those in the estimated stock network, $FQ_j [Net_{L}^{O} \bigcap
Net_{L}^{E}]$. This is the case among the number of $M$ stocks
with more than degree, $k=1,2,\cdots, Max$, in the original stock
network [13, 16]. Here, the survivor ratio can vary between $0.0
\leq r_s \leq 1.0$. If $r_s=0$, two stock networks have a
completely different structure; however, if $r_s=1$, they have the
same structure. We calculated the survival ratio according to the
various number of links tp other stocks in the original stock
network, $L \geq k$. Thus, the observation that there is an
increase in the number of links makes it possible to confirm
whether stocks having a number of links different from those of
other stocks are influenced differently by correlations having
various eigenvalue properties.

\section{Results}
\label{sec:RESULTS}

In this section, we present the observed results for stock network
properties using the RMT method. Using the $N$ individual stocks
traded on the stock market index of each country, we estimated the
stock network by utilizing the MST method. We estimated the
correlation matrix, $C_{i=1}^{S(i)}$, reflecting various
eigenvalue properties as well as the individual stock numbers of
each country. In addition, we investigated whether the largest
eigenvalue property estimated by the RMT plays an important role
in terms of the formation of a stock network. That is, we tested
the original stock network and compared it to the estimated stock
network from the correlation matrix reflecting all the eigenvalue
properties except the largest eigenvalue. In order to calculate
the degree of consistency between the original stock network and
the estimated stock network, we used the survivor ratio, $r_s$.

The results are presented in Fig. 1. In Fig. 1, the x-axis
indicates the number of eigenvalues. The axis x denotes the
normalized number of eigenvalues,
$L^{*}=(L_{i}-L_{Min})/(L_{Max}-L_{Min})-1$, from -1 to +1 using
the minimum, $L_{Min}$, and maximum, $L_{Max}$, of the number of
eigenvalues of each country. This is because the number of
eigenvalues for each country is different. Axis y represents the
average of the survivor ratio. Fig. 1(a) shows that the survivor
ratio of stocks having a number of links to other stocks is more
than one, $L \geq 1$. It represents a degree of consistency for
all stocks which exist in a network. Fig. 1(d) is the survivor
ratio of stocks having the largest number of links, $L \geq Max$;
therefore, it shows a degree of consistency for stocks having the
greatest number of links in a network. Figs. 1(c) and 1(d) are the
same as Figs. 1(a) and 1(b), respectively. The difference is that
the fixed starting point of the eigenvalue used to estimate the
correlation matrix is the second largest eigenvalue,
$\lambda_{i=2}$. That is, we did not include the largest
eigenvalue property. The circles (red), triangles (magenta),
diamonds (cyan), stars (blue), squares (green), and pentagrams
(black) indicate Korea, Japan, Canada, the USA, Italy, and the UK,
respectively.

\begin{figure}[tb]

\includegraphics[height=10cm, width=10cm]{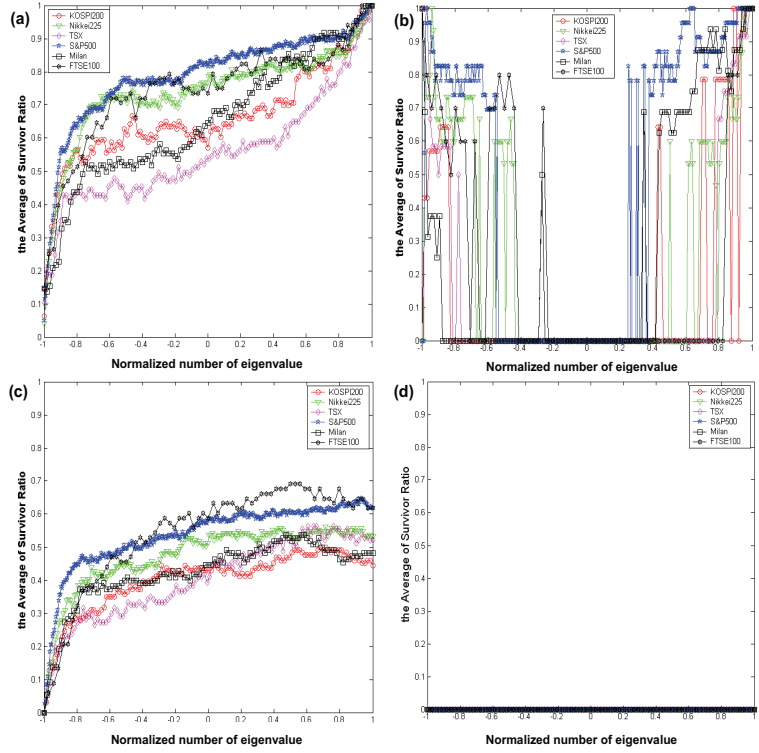}

\caption[0]{Figs. 1(a) and (b) show the survivor ratio between the
estimated stock network from the RMT and the original stock
network. (a) shows that the survivor ratio of stocks having a
number of links to other stocks is greater than one. (b) is the
survivor ratio of stocks having the largest number of links. Figs.
1(c) and (d) are the same as Figs. 1(a) and (b), respectively. The
difference is that the fixed starting point of the eigenvalue used
to estimate the correlation matrix is the second largest
eigenvalue. That is, we did not include the largest eigenvalue
property. We normalized the number of eigenvalues from -1 to +1
using the maximum and minimum number of eigenvalues of each
country, because the number of eigenvalues for each country is
different. The circles (red), triangles (magenta), diamonds
(cyan), stars (blue), squares (green), and pentagrams (black)
indicate Korea, Japan, Canada, the USA, Italy, and the UK,
respectively.}

\end{figure}

According to the results, regardless of the data, as the
correlation matrix reflects the more eigenvalue property, the
estimated stock network from the correlation matrix gradually
increases the degree of consistency with the original stock
network [Fig. 1a]. The interesting point is that in Fig. 1(b),
stocks having the largest number of links show a significant
change before and after the eigenvalue range, $\lambda_{-}^{RM}
\leq \lambda_{i} \leq \lambda_{+}^{RM}$. This is predicted by the
RMT, where $\lambda_{+}^{RM}$ and $\lambda_{-}^{RM}$ are maximum
and minimum eigenvalues of the RMT, respectively. That is, when
the correlation matrix includes the eigenvalue properties
deviating from the range, $\lambda_{+}^{RM} < \lambda_{i} \leq
\lambda_{i=1}$, the survivor ratio increases rapidly.
Significantly, the survivor ratio decreases as the correlation
matrix reflects the more eigenvalue property within the ranges,
while the survivor ratio increases again as the correlation matrix
reflects the more eigenvalue property within the range,
$\lambda_{i=N} \leq \lambda_{i} \ < \lambda_{-}^{RM}$. These
results suggest that each stock with a different number of links
to other stocks in the original stock network exhibits different
behavior in terms of the correlation matrix reflecting various
eigenvalue properties. In Fig. 1(c), although the correlation
matrix reflects all the eigenvalue properties except the largest
eigenvalue, the estimated stock network can explain $40 \sim 60\%$
of the interactions between stocks with the original stock
network. In particular, in Fig. 1(d), stocks having the largest
number of links directly influence the largest eigenvalue property
because the survivor ratio is 0. These results suggest that the
largest eigenvalue, which reveals the market index as a
representative single factor in the financial literature, is a
significant deterministic factor in terms of the formation of a
stock network. However, we also found that this property alone
cannot sufficiently explain the interaction between stocks.

\section{Conclusions}
\label{sec:CONCLUSIONS}

We investigated the topological properties of stock networks
through a comparison of the original stock networks with the
estimated stock networks using a correlation matrix having various
eigenvalue properties calculated by the RMT method. We used the
individual stocks traded on the representative stock market
indices of Korea, Japan, Canada, the USA, Italy, and the UK,
respectively. Regardless of the data of the countries involved, as
the correlation matrix reflects the more eigenvalue property, the
estimated stock network from the correlation matrix gradually
increases the degree of consistency with the original stock
network. The interesting point is that each stock with a different
number of links to other stocks in the original stock network
shows a different response before and after the range of
eigenvalues predicted by the RMT. In particular, when we did not
include the property of the largest eigenvalue in the correlation
matrix estimated by RMT, although it reflected all eigenvalue
properties except the largest eigenvalue, the estimated stock
network could not adequately explain the formation of an original
stock network. Through these results, we have demonstrated that
the largest eigenvalue, which reveals the market index as a
representative single factor in the field of finance, must be a
significant deterministic factor in terms of the formation of a
stock network. However, the properties of the largest eigenvalue
alone cannot fully explain the interaction between stocks.

\end{document}